\documentstyle[pre,aps]{revtex}
\begin{document}
\draft
\title{Non-Abelian linear Boltzmann equation and quantum correction to
  Kramers and Smoluchowski equation}
\author{Bassano~Vacchini}
\address{Dipartimento di Fisica
dell'Universit\`a di Milano and INFN, Sezione di Milano,
Via Celoria 16, I-20133 Milan,
Italy}
\date{\today}
\maketitle
\begin{abstract}
  A quantum linear Boltzmann equation is proposed, constructed in
  terms of the operator-valued dynamic structure factor of the
  macroscopic system the test particle is interacting with. Due to
  this operator structure it is a non-Abelian linear Boltzmann
  equation and when expressed through the Wigner function it allows
  for a direct comparison with the classical one. Considering a
  Brownian particle the corresponding Fokker-Planck equation is
  obtained in a most direct way taking the limit of small energy and
  momentum transfer. A typically quantum correction to the Kramers
  equation thus appears, describing diffusion in position and further
  implying a correction to Einstein's diffusion coefficient in the
  high temperature and friction limit in which Smoluchowski equation
  emerges.
\end{abstract}
\pacs{05.30.-d, 05.40.Jc, 05.60.Gg, 03.65.Yz} The study of
irreversible processes, even in relatively simple cases in which one
considers the reduced dynamics of a system with few degrees of freedom
interacting with a suitable reservoir, is a subject of major interest
in both classical and quantum mechanics and a necessity whenever one
wants to cope with real experiments, possibly starting from the
underlying microphysical dynamics. In particular in quantum mechanics
irreversible incoherent, and possibly dissipative, processes play a
decisive role in determining the feasibility of quantum
computers~\cite{ZeilingerQC} and under the unifying concept of
decoherence might even have a deep impact in the understanding of
quantum mechanics and of its relationship with the classical
description of the world~\cite{Kiefer}, provided a precise connection
with microphysical models can be given.
\par
In the specific case of a microsystem whose interaction with a
macroscopic system or reservoir can be described in terms of
collisions, a major tool for the classical description of its dynamics
is the linear Boltzmann equation or transport equation. This equation
has proved essential in fields like neutron transport
theory~\cite{Williams} and is still the standard reference for this
kind of dynamics even at quantum level~\cite{Timmermans-Ferrari}, the
only improvement lying in the introduction of the quantum collision
cross-section in place of the classical one. This kinetic equation,
provided a suitable expansion of the collision term can be performed,
then leads to a Fokker-Planck equation for the study of
dissipation~\cite{Risken}, which for Brownian motion is the so-called
Kramers equation. Considering situations in which dissipation is very
important, Kramers equation in turn leads to the so-called
Smoluchowski equation in which Einstein's diffusion coefficient
appears.
\par
In this paper we give the structure of a fully quantum, non-Abelian
linear Boltzmann equation, non-Abelian just due to the presence in it
of the operator-valued rather than c-number dynamic structure factor
of the macroscopic system, as typical of the quantum realm, giving a
physical example of a recently obtained mathematical result on
Lindblad generators of translation-covariant completely positive
quantum dynamical semigroups~\cite{HolevoRMP32-HolevoRMP33-HolevoJMP}.
Considering the Brownian motion of a test particle in a gas a most
straightforward expansion of the non-Abelian linear Boltzmann
equation, written in terms of the Wigner function, leads to a Kramers
equation with a typical quantum correction linked to position
diffusion vanishing in the semiclassical limit $\hbar\rightarrow 0$.
Such a correction ensures positivity at quantum level and is connected
to the time evolution of the off-diagonal matrix elements of the
statistical operator in the momentum representation, which can only be
given in a fully quantum mechanical version of the linear Boltzmann
equation. Studying the high friction limit of this Fokker-Planck
equation analogously to the classical case, in the same spirit, also a
quantum correction to Einstein's diffusion coefficient appears.
\section{Non-Abelian linear Boltzmann equation}
Let us start writing the linear Boltzmann equation for the classical
case, still relying on the quantum expression for the collision
cross-section, in a way which is actually not the most common one.
According to a famous paper by van Hove~\cite{vanHove}, the
energy-dependent differential cross-section per target particle
describing scattering of a microscopic probe off a macroscopic sample
is given by
\begin{equation}
  \label{diff}
  \frac{d^2 \sigma}{d\Omega_{p'} dE_{p'}}  =
\frac{p'}{p}\Sigma (q)
  S ({\bbox{q}},E)
\end{equation}
if the momentum of the microscopic probe changes from ${\bbox{p}}$ to
${\bbox{p}}' = {\bbox{p}}+{\bbox{q}}$.  $\Sigma (q)$ is here the
collision cross-section for the single scattering event, given by
$\Sigma (q)=M^2 (2\pi\hbar)^3\frac{2\pi}{\hbar}| \tilde{t} (q) |^2$,
with $\tilde{t} (q)$ Fourier transform of the T-matrix describing the
one-pair collisions between the test particle and the particles
constituting the sample, supposed to depend only on the modulus of the
momentum transfer ${\bbox{q}}$. $S ({\bbox{q}},E)$ is a two-point
correlation function known as dynamic structure factor~\cite{Lovesey},
reflecting the equilibrium many-body properties of the fluid,
depending on momentum ${\bbox{q}}$ and energy
$E=E({\bbox{q}},{\bbox{p}})={ q^2 \over 2M } + { {\bbox{p}} \cdot
  {\bbox{q}} \over M }$ transferred to the test particle of mass $M$,
whose general expression is given by the Fourier transform with
respect to energy and momentum transfer of the time dependent spatial
autocorrelation function according to:
\begin{eqnarray*}
  {S} ({\bbox{q}},E)&=&
        {  
        1  
        \over  
         2\pi\hbar
        }  
        \int dt 
        {\int d^3 \! {\bbox{x}} \,}        
        e^{
        {
        i
        \over
         \hbar
        }
        (E t -
        {\bbox{q}}\cdot{\bbox{x}})
        }  
      \\
\nonumber
&&
\hphantom{        {  
        1  
        \over  
         2\pi\hbar
        }  
        \int dt 
}
\times
        {
        1
        \over
         N
        }
        {\int d^3 \! {\bbox{y}} \,}
        \left \langle  
         N({\bbox{y}})  
         N({\bbox{x}}+{\bbox{y}},t)
         \right \rangle.
\end{eqnarray*}
Exploiting (\ref{diff}) the linear Boltzmann equation in the absence
of external potentials may be written
as~\cite{Williams}
\begin{eqnarray}
  \label{lbe}
  && \frac{\partial}{\partial t}f ({\bbox{x}},{\bbox{p}},t)=
  -\frac{{\bbox{p}}}{M}\cdot {\mathbf \nabla}_x 
  f ({\bbox{x}},{\bbox{p}},t)
 +\frac{n}{M^2}
        \int d^3\!
        {{\bbox{q}}}
        \,  
        \Sigma (q)
  \\
 &&
 \hphantom{ju}
 \times
        \left[ S({\bbox{q}},{ {\bbox{p}}}-{\bbox{q}})f
          ({\bbox{x}},{\bbox{p}}-{\bbox{q}},t)-S({\bbox{q}},
          {\bbox{p}})f ({\bbox{x}},{\bbox{p}},t)\right] 
        \nonumber
\end{eqnarray}
with $S({\bbox{q}},{\bbox{p}})\equiv S ({\bbox{q}},E)$ the dynamic
structure factor of the homogeneous macroscopic fluid of density $n$
the particle is interacting with. So far the classical case, all
expressions appearing in (\ref{lbe}) being c-number functions.
\par
Recent work on the study of subdynamics in non-relativistic quantum
field theory~\cite{art6} has led to a scattering theory derivation of
a particular structure of master-equation~\cite{art3,reply,art4},
which goes beyond the usual limitation of linear coupling and is a
natural candidate to be considered as a fully quantum linear Boltzmann
equation, as we will show in the sequel. Instead of an equation for
the classical distribution function it is an equation for the full
statistical operator ${\hat \rho}$ associated to the test particle,
describing not only its momentum distribution, but also its coherence
properties. These are particularly relevant for the time evolution of
the off-diagonal matrix elements, related to the phenomenon of
decoherence, and for the low temperature behavior, where the classical
particle viewpoint is insufficient and wave effects are to be taken
into account. The quantum kinetic equation is given by:
\begin{eqnarray}
  \label{nalbe}
  &&   
  {  
        d {\hat \rho}  
        \over  
                      dt
        }  
        =
        -
        {i \over \hbar}
        [
        {\hat {{\sf H}}}_0
        ,
        {\hat \rho}
        ]
                +
        {n \over M^2}
        \int d^3\!
        {{\bbox{q}}}
        \,  
        \Sigma (q)
  \\
 &&
 \hphantom{ju}
 \times
      \Biggl[
        e^{{i\over\hbar}{{\bbox{q}}}\cdot{\hat {{\sf x}}}}
        \sqrt{
        S({{\bbox{q}}},{\hat {{\sf p}}})
        }
        {\hat \rho}
        \sqrt{
        S({{\bbox{q}}},{\hat {{\sf p}}})
        }
        e^{-{i\over\hbar}{{\bbox{q}}}\cdot{\hat {{\sf x}}}}
        -
        \frac 12
        \left \{
        S({{\bbox{q}}},{\hat {{\sf p}}}),
        {\hat \rho}
        \right \}
        \Biggr],
        \nonumber
\end{eqnarray}
with ${\hat {{\sf x}}}$ and ${\hat {{\sf p}}}$ position and momentum
operator for the test particle, ${\hat {{\sf H}}}_0={\hat {{\sf p}}}^2
/ 2M$, $S({{\bbox{q}}},{\hat {{\sf p}}})$ the operator-valued dynamic
structure factor, which is a positive operator due to (\ref{diff}).
Equation (\ref{nalbe}) addresses the same physics as the linear
Boltzmann equation but is non-Abelian due to the appearance in it of
operators instead of c-number functions and it provides a physical
realization of the general mathematical structure of
translation-covariant quantum dynamical
semigroup~\cite{HolevoRMP32-HolevoRMP33-HolevoJMP}, other known
physical examples restricting to the diffusive case. Neglecting the
dependence on the momentum operator ${\hat {{\sf p}}}$ and therefore
on the energy transfer of the dynamic structure factor in
(\ref{nalbe}), so that its appearance only corrects the integration
measure, one recovers a structure of master-equation typically
proposed on a more phenomenological basis for the study of the
phenomenon of decoherence, which has gone through many
refinements~\cite{Joos-Zeh-Gallis90-AlickiPRA}, all missing the
correct description of momentum and energy transfer, which may not be
critical for the description of short-time decoherence but are crucial
for the approach to the correct stationary state. In fact while the
models in~\cite{Joos-Zeh-Gallis90-AlickiPRA} predict a steady growth
of kinetic energy and do not admit a stationary state, provided the
fluid is in a $\beta$-KMS state or equivalently the dynamic structure
factor satisfies the detailed balance condition, eq.~(\ref{nalbe})
admits a stationary solution of the form $ \exp ({- \beta { {\hat
      {{\sf p}}}^2 / 2M } })$, with $M$ mass of the test particle and
$\beta$ inverse temperature of the fluid~\cite{art5}. In order to
clarify analogies and differences between (\ref{lbe}) and
(\ref{nalbe}) we will now consider a dynamics in which als the
off-diagonal elements of the statistical operator are of relevance,
corresponding to position diffusion, focusing on a system in which the
dynamic structure factor can be explicitly calculated, a free gas of
particles of mass $m$ satisfying Maxwell-Boltzmann statistics (for the
extension to quantum statistics see~\cite{art5}), the expression of
the correlation function being in this case incidentally the same for
both classical and quantum realm~\cite{Lovesey}
\begin{equation}
  \label{7}
        S_{\rm \scriptscriptstyle MB}({\bbox{q}},{\bbox{p}})
        =
        \sqrt{\frac{\beta m}{2\pi}}        
        {
        1
        \over
        q
        }
       e^{-{
        \beta
        \over
             8m
        }
        {
        (2mE({\bbox{q}},{\bbox{p}}) + q^2)^2
        \over
                  q^2
        }}
        .
\end{equation}
\section{Quantum correction to Kramers equation}
As a first step we recover from (\ref{lbe}) the classical Kramers
equation for the Brownian motion of a massive test particle in a gas
of much lighter ones. Calling $\alpha=m/M$ the ratio between the
masses and considering the Brownian limit of small $\alpha$,
corresponding to small energy transfer, the correct limiting
expression of (\ref{7}), still satisfying the detailed balance
condition is~\cite{art5}
\begin{equation}
        \label{22}
        S^{\scriptscriptstyle\infty}_{\rm \scriptscriptstyle
          MB}({\bbox{q}},{\bbox{p}}) 
        =
        \sqrt{\frac{\beta m}{2\pi}}
        {
        1
        \over
        q
        }
        e^{
        -{
        \beta
        \over
             8m
        }
        q^2
        }
        e^{
        -\frac{\beta}{2}
        E({\bbox{q}},{\bbox{p}})
        }.
\end{equation}
Starting from the linear Boltzmann equation written in terms of the
dynamic structure factor as in (\ref{lbe}) and exploiting (\ref{22})
one obtains in a most straightforward way Kramers equation, a point
usually neglected in textbooks~\cite{Risken}. Substituting (\ref{22})
in (\ref{lbe}) one in fact has
\begin{eqnarray}
  \label{cl}
\frac{\partial}{\partial t}f ({\bbox{x}},{\bbox{p}},t)&=& 
  -\frac{{\bbox{p}}}{M}\cdot {\mathbf \nabla}_x 
  f ({\bbox{x}},{\bbox{p}},t)
\\
\nonumber
&\hphantom{=}&
 +\frac{n}{M^2}
        \sqrt{\frac{\beta m}{2\pi}}
    \int d^3\!
        {{\bbox{q}}}
        \,  
        {
        \Sigma (q)
        \over
        q
        }
        e^{-
        {
        \beta
        \over
             8m
        } (1+2\alpha)
        {{{q}}^2}
        }
        \left[ e^{-{\bbox{q}}\cdot {\mathbf \nabla}_p}
         -1
          \right]
          e^{-{\beta\over 2M}{\bbox{q}}\cdot {\bbox{p}}}
          f({\bbox{x}},{\bbox{p}},t) 
\end{eqnarray}
and considering the limit of small momentum transfer ${\bbox{q}}$,
expanding the exponentials and keeping terms up to second order one
immediately has for an isotropic medium ($ {\bbox{q}}^2_i = \frac 13
q^2 $)
\begin{eqnarray}
   \frac{\partial}{\partial t}f ({\bbox{x}},{\bbox{p}},t) &=&
  -\frac{{\bbox{p}}}{M}\cdot {\mathbf \nabla}_x 
  f ({\bbox{x}},{\bbox{p}},t)
\\
\nonumber
&\hphantom{=}&
 +\frac{n\beta}{6M^3}
        \sqrt{\frac{\beta m}{2\pi}}
       \int d^3\!
        {{\bbox{q}}}
        \,  
        q        \Sigma (q)
        e^{-
        {
        \beta
        \over
             8m
        } (1+2\alpha)
        {{{q}}^2}
        }
      \sum_{i=1}^3 \frac{\partial}{\partial p_i}
      \left[ \left( p_i\!+\! \frac{M}{\beta}\frac{\partial}{\partial
            p_i}\right) f({\bbox{x}},{\bbox{p}},t) 
          \right]
\end{eqnarray}
and therefore Kramers equation for the description of Brownian motion
\begin{equation}
  \label{K}
   \frac{\partial}{\partial t}f ({\bbox{x}},{\bbox{p}},t)= 
  -\frac{{\bbox{p}}}{M}\cdot {\mathbf \nabla}_x 
  f ({\bbox{x}},{\bbox{p}},t)
 +\eta
      \left[{\mathbf \nabla}_p \cdot
        ({\bbox{p}}f({\bbox{x}},{\bbox{p}},t)) 
        + \frac{M}{\beta}{\bf \Delta}_p f({\bbox{x}},{\bbox{p}},t) 
          \right],
\end{equation}
with a friction coefficient explicitly given by 
\begin{displaymath}
\eta=\frac{1}{6}
\frac{n}{M^3}\beta
\sqrt{\frac{\beta m}{2\pi}} \int d^3\!  {{\bbox{q}}} \, q \Sigma
(q) 
e^{-\frac{\beta}{8m}(1+2\alpha){{{q}}^2}}.
\end{displaymath}
\par
We now want to obtain the corresponding result from the non-Abelian
linear Boltzmann equation (\ref{nalbe}). To do this it is particularly
convenient to use the Wigner function, which even if it is not a
well-defined probability density, well serves the purpose of comparing
quantum versus classical equations. Introducing the Wigner function
corresponding to the statistical operator $\hat{\rho}$ by
\begin{displaymath}
f^{\rm \scriptscriptstyle
  W}_{\rho} ({\bbox{x}},{\bbox{p}},t)= \int \frac{d^3\!
  {\bbox{k}}}{(2\pi\hbar)^3} \, 
e^{\frac{i}{\hbar}{\bbox{x}} \cdot{\bbox{k}}}
\langle{\bbox{p}}
+{{\bbox{k}}}/{2}|\hat{\rho}|{\bbox{p}}-{{\bbox{k}}}/{2}\rangle,
\end{displaymath}
eq.~(\ref{nalbe}) may be easily rewritten for a Brownian particle in a
Maxwell-Boltzmann gas, exploiting (\ref{22}), as
\begin{eqnarray}
  \label{q}
   \frac{\partial}{\partial t}f^{\rm \scriptscriptstyle
     W}_{\rho} ({\bbox{x}},{\bbox{p}},t)&=& 
  -\frac{{\bbox{p}}}{M}\cdot {\mathbf \nabla}_x 
  f^{\rm \scriptscriptstyle
     W}_{\rho} ({\bbox{x}},{\bbox{p}},t)
  \\
 &\hphantom{=}&
 \nonumber
 +\frac{n}{M^2}
 \sqrt{\frac{\beta m}{2\pi}}
    \int d^3\!
        {{\bbox{q}}}
        \,  
        {
        \Sigma (q)
        \over
        q
        }
        e^{-
        {
        \beta
        \over
             8m
        } (1+2\alpha)
        {{{q}}^2}
        }
\\
\nonumber
&\hphantom{=}&
\hphantom{\frac{n}{M^2}
 \sqrt{\frac{\beta m}{2\pi}}
    \int d^3\!
        {{\bbox{q}}}
        \,  
        {
        \Sigma (q)
        \over
        q
        }}
\times
        \left[ e^{-{\bbox{q}}\cdot {\mathbf \nabla}_p}
         - \cos \left(\frac{\beta\hbar}{4M}{\bbox{q}}\cdot{\mathbf
             \nabla}_x\right) 
          \right]
          e^{-{\beta\over 2M}{\bbox{q}}\cdot {\bbox{p}}}
          f^{\rm \scriptscriptstyle
     W}_{\rho}({\bbox{x}},{\bbox{p}},t) .
\end{eqnarray}
The comparison between (\ref{cl}) and (\ref{q}) is straightforward:
the cosine term in (\ref{q}), arising from the non-Abelian structure
of (\ref{nalbe}), is replaced by a factor one in the classical case,
and this result can be simply obtained taking the semiclassical limit
$\hbar\rightarrow 0$ in (\ref{q}), the argument of the cosine
depending linearly on $\hbar$. The quantum correction to Kramers
equation then comes from this contribution and amounts to a term
corresponding to diffusion in position, as can immediately be seen
considering the small momentum transfer limit of (\ref{q})
\begin{eqnarray}
  \label{Kq}
   \frac{\partial}{\partial t}f^{\rm \scriptscriptstyle
     W}_{\rho} ({\bbox{x}},{\bbox{p}},t)&=&
  -\frac{{\bbox{p}}}{M}\cdot {\mathbf \nabla}_x 
  f^{\rm \scriptscriptstyle
     W}_{\rho} ({\bbox{x}},{\bbox{p}},t)
  \\
 &\hphantom{=}&
 \nonumber
 +\eta
      \left[{\mathbf \nabla}_p \cdot
        ({\bbox{p}}f^{\rm \scriptscriptstyle
     W}_{\rho}({\bbox{x}},{\bbox{p}},t)) 
        + \frac{M}{\beta}{\bf \Delta}_p f^{\rm \scriptscriptstyle
     W}_{\rho}({\bbox{x}},{\bbox{p}},t) 
  + \frac{\beta}{M}\frac{\hbar^2}{16}{\bf \Delta}_x f^{\rm
    \scriptscriptstyle 
     W}_{\rho}({\bbox{x}},{\bbox{p}},t)         \right],
\end{eqnarray}
with $\eta$ as in (\ref{K}). Thus position and momentum diffusion do
appear together in the quantum description of Brownian motion, even
though with different weights, as appears from the different $\beta$
dependence of the two contributions, the additional diffusion term
vanishing in the limit $\hbar\rightarrow 0$. This can be seen more
clearly introducing the thermal momentum spread ${\Delta p}^2_{\rm
  \scriptscriptstyle th}=\frac{M}{\beta}$ and the thermal position
spread or square thermal wavelength ${\Delta x}^2_{\rm
  \scriptscriptstyle th}=\frac{\beta\hbar^2}{4M}$, satisfying the
minimum uncertainty relation ${\Delta p}_{\rm \scriptscriptstyle
  th}{\Delta x}_{\rm \scriptscriptstyle th}=\frac{\hbar}{2}$, so that
one has
\begin{eqnarray}
  \label{th}
   \frac{\partial}{\partial t}f^{\rm \scriptscriptstyle
     W}_{\rho} 
&=&
  -\frac{{\bbox{p}}}{M}\cdot {\mathbf \nabla}_x 
  f^{\rm \scriptscriptstyle
     W}_{\rho} 
 +\eta
      \left[{\mathbf \nabla}_p \cdot
        ({\bbox{p}}f^{\rm \scriptscriptstyle
     W}_{\rho})
   \hphantom{\frac 14}
   \right.
\\
\nonumber
&&
\hphantom{pippo}
\left.
        + {\Delta p}^2_{\rm \scriptscriptstyle th}{\bf \Delta}_p
        f^{\rm \scriptscriptstyle 
     W}_{\rho}
  + \frac 14 {\Delta x}^2_{\rm \scriptscriptstyle th}{\bf \Delta}_x
  f^{\rm \scriptscriptstyle 
     W}_{\rho}
 \right].
\end{eqnarray}
\par
The Fokker-Planck equation (\ref{Kq}) or (\ref{th}) can be recast in
operator form, leading to the well-known structure~\cite{Isar99}
\begin{eqnarray}
        \label{qth}
        {  
        d {\hat \rho}  
        \over  
                dt  
        }  
        =  
        &-&
        {i\over\hbar}
        [
        {{\hat {{\sf H}}}_0}
        ,{\hat \rho}
        ]
        -
        {i\over\hbar}
        \frac{\eta}{2}
        \sum_{i=1}^3
        \left[  
        {\hat {{\sf x}}}_i ,
        \left \{  
        {\hat {\sf p}}_i,{\hat \rho}
        \right \}  
        \right]            
\\
        &-&
        {
        \eta
        \over
         \hbar^2
        }{\Delta p}^2_{\rm \scriptscriptstyle th}
        \sum_{i=1}^3
        \left[  
        {\hat {\sf x}}_i,
        \left[  
        {\hat {\sf x}}_i,{\hat \rho}
        \right]  
        \right]  
        -{
        \eta
        \over
         \hbar^2
        }\frac 14 {\Delta x}^2_{\rm \scriptscriptstyle th}
        \sum_{i=1}^3
        \left[  
        {\hat {{\sf p}}}_i,
        \left[  
        {\hat {{\sf p}}}_i,{\hat \rho}
        \right]  
        \right]  
  ,
\nonumber
\end{eqnarray}
which could have been obtained directly from (\ref{nalbe}) as
in~\cite{art5}, without going through the use of the Wigner function,
missing however in this way the particularly manifest appearance of
the quantum correction which provides a inhomogeneous contribution to
the collision term in the linear Boltzmann equation through the
appearance of the cosine in (\ref{q}).  The appearance of both
position and momentum diffusion in (\ref{th}) or equivalently
(\ref{qth}) is due to the fact that (\ref{nalbe}) is an evolution
equation not only for the diagonal elements of the statistical
operator giving the momentum distribution of the test particle, but
also for its quantum coherence properties, which actually require
preservation of the positivity of the statistical operator. In fact
the quantum correction in (\ref{th}), actually corresponding to the
double commutator with the momentum operator in (\ref{qth}), is
necessary for (\ref{qth}) to have a Lindblad structure ensuring
complete positivity (actually corresponding with positivity for
Fokker-Planck equations of the form (\ref{qth})~\cite{Talkner}).
Correctness and relevance of this term are heavily
debated~\cite{debate,art3,reply,art5}, also in connection with the
meaning of complete positivity~\cite{cp}, and its appearance as a
typical quantum correction coming from the Brownian and small momentum
transfer limit of the non-Abelian linear Boltzmann equation
(\ref{nalbe}) expressed in terms of the operator-valued dynamic
structure factor sheds light on its physical origin. The quantum
Fokker-Planck equation (\ref{qth}) due to the minimum uncertainty
relation ${\Delta p}_{\rm \scriptscriptstyle th}{\Delta x}_{\rm
  \scriptscriptstyle th}=\frac{\hbar}{2}$ has furthermore the
distinguishing feature of being expressible in terms of a single
generator for Cartesian direction~\cite{art3}, a property actually
required for the diffusive component of the generator of a
translation-covariant
semigroup~\cite{HolevoRMP32-HolevoRMP33-HolevoJMP}: both (\ref{nalbe})
and (\ref{qth}) are proper generators of completely positive quantum
dynamical semigroups, admitting the correct equipartition stationary
solution and being invariant under translations. The three features of
complete positivity, equipartition and covariance are in fact not
contradictory provided one considers covariance under the symmetry
relevant to the physical problem, which is not necessarily translation
invariance~\cite{Tannor97}, otherwise, independently on positivity or
complete positivity of the time evolution, covariance would lead to
high non-uniqueness of the stationary solution~\cite{art8}.  Equations
(\ref{nalbe}) and (\ref{qth}) are essentially the two possible
structures of generators of translation-covariant semigroups, the
results exposed in~\cite{HolevoRMP32-HolevoRMP33-HolevoJMP} giving the
quantum counterpart of the well-known classical result according to
which the only meaningful approximation of the integro-differential
linear Boltzmann equation as a differential equation is a second order
Fokker-Planck equation~\cite{Pawula}.
\par
Now that the quantum correction to Kramers equation has been given,
which unlike previous results is not dependent on the presence of
external potentials~\cite{Ishioka}, but only relates to Heisenberg's
uncertainty principle~\cite{Isar99}, we move on to investigate whether
this correction also has consequences on the Smoluchowski equation for
the description of Brownian motion in the limit of high temperature
and friction, when momentum is expected to quickly relax to its
equilibrium value and one is only interested in the time evolution of
the slowly varying marginal position distribution
\begin{displaymath}
\sigma^{\rm
  \scriptscriptstyle W}_{\rho}({\bbox{x}},t)= \int d^3\!  {{\bbox{p}}}
\,f^{\rm \scriptscriptstyle W}_{\rho}({\bbox{x}},{\bbox{p}},t).
\end{displaymath}
\section{Quantum correction to Smoluchowski equation}
We therefore start from (\ref{Kq}), essentially following the
derivation of the Smoluchowski equation from the Kramers equation in
the high friction limit given by van Kampen~\cite{vanKampen}, though
more refined derivations have been given~\cite{Bocquet}, differing
however only for higher order contributions. Setting
\begin{equation}
  \label{coeff}
  \eta\frac{\beta}{M}\frac{\hbar^2}{16}=\frac{\eta}{4}{\Delta
    x}^2_{\rm \scriptscriptstyle th}\equiv D_{xx}
\end{equation}
and having in mind a high temperature, strong friction situation in
which the quantum position diffusion term can be considered as a
perturbation, we write (\ref{Kq}) as
\begin{eqnarray}
  \label{100}
&&
  {\mathbf \nabla}_p \cdot
        ({\bbox{p}}f^{\rm \scriptscriptstyle
     W}_{\rho})
        + \frac{M}{\beta}{\bf \Delta}_p f^{\rm \scriptscriptstyle
     W}_{\rho}
   =
\\
\nonumber
&&
\hphantom{pippo}
   \frac{1}{\eta}
\left[\frac{\partial}{\partial t}f^{\rm \scriptscriptstyle
     W}_{\rho} 
   +\frac{{\bbox{p}}}{M}\cdot {\mathbf \nabla}_x 
  f^{\rm \scriptscriptstyle
    W}_{\rho} 
  - D_{xx}{\bf \Delta}_x f^{\rm \scriptscriptstyle
     W}_{\rho}
\right].
\end{eqnarray}
Introducing the differential operators
\begin{eqnarray*}
  K^{(0)}\left[g\right]&=& {\mathbf \nabla}_p \cdot
        ({\bbox{p}}g) 
        + \frac{M}{\beta}{\bf \Delta}_p g
   \\
   \frac{1}{\eta}L^{(1)}\left[g\right]&=&
   \frac{1}{\eta}
\left[\frac{\partial}{\partial t}g
   +\frac{{\bbox{p}}}{M}\cdot {\mathbf \nabla}_x 
  g  - D_{xx}{\bf \Delta}_x g
 \right]
\end{eqnarray*}
and expressing $f^{\rm \scriptscriptstyle
     W}_{\rho}$ as an expansion in powers of
${1}/{\eta}$
\begin{equation}
  \label{103}
  f^{\rm \scriptscriptstyle
     W}_{\rho}
   =
   f^{\rm \scriptscriptstyle
     W}_{\rho}{}^{(0)}+\frac{1}{\eta}f^{\rm
     \scriptscriptstyle W}_{\rho}{}^{(1)}+\ldots ,
\end{equation}
eq.~(\ref{100}) may be rewritten as $ K^{(0)}\left[f^{\rm
    \scriptscriptstyle W}_{\rho}\right]=
\frac{1}{\eta}L^{(1)}\left[f^{\rm \scriptscriptstyle W}_{\rho}\right]
$ and solved by iteration using (\ref{103}) and equating on both sides
contributions of the same order in $\eta$. The only difference with
respect to the classical situation considered in~\cite{vanKampen} lies
in the structure of the $L^{(1)}$ operator, where also the Laplacian
with respect to position appears. Going through the very same
procedure, integrating over the fast degree of freedom given by the
momentum dependence of the Wigner distribution function, and
neglecting terms higher than first order in $\frac{1}{\eta}$ one
obtains again for the marginal position distribution $\sigma^{\rm
  \scriptscriptstyle W}_{\rho}({\bbox{x}},t)$ the Smoluchowski
equation
\begin{equation}
  \label{Sq}
  \frac{\partial}{\partial t}\sigma^{\rm \scriptscriptstyle
     W}_{\rho} ({\bbox{x}},t)=
 \left( \frac{1}{\eta M\beta}+ D_{xx}\right)
   {\bf \Delta}_x \sigma^{\rm \scriptscriptstyle
     W}_{\rho}({\bbox{x}},t),
\end{equation}
with a diffusion coefficient which is however not simply Einstein's,
but has a quantum correction, vanishing in the semiclassical limit
$\hbar\rightarrow 0$, just given by the coefficient (\ref{coeff})
responsible for position diffusion in the quantum Kramers equation
(\ref{Kq}). The new overall coefficient can be written as $1 / (\eta
M\beta) (1+ (\eta\beta\hbar)^2/16)$, so that the correction is
actually given by the square ratio between two characteristic times,
$\beta\hbar$ for the bath and $1/\eta$ for the test particle, which
should be much less than one according to the Markov approximation.
\par
We have thus proposed a fully quantum, non-Abelian linear Boltzmann
equation given by (\ref{nalbe}) and expressed in terms of the T-matrix
describing collisions and the operator-valued dynamic structure factor
of the environment, thus going beyond the usual restriction of linear
dissipation. Considering a massive test particle interacting with a
gas of much lighter ones the linear Boltzmann equation written in
terms of the dynamic structure factor leads in a most straightforward
way to the Kramers equation for the description of Brownian motion in
both classical and quantum case.  One simply has to consider the
clear-cut physical limit in which momentum transfer and energy
transfer (or equivalently ratio between the masses) are small, thus
obtaining in the classical case Kramers equation and in the quantum
case (\ref{Kq}) with a peculiar, heavily debated quantum correction
linked to position diffusion, whose appearance is to be traced back to
the $\hbar$ dependent, inhomogeneous cosine correction in (\ref{q})
vanishing in the semiclassical limit. Further studying the case of
high temperature and friction the quantum Smoluchowski equation
(\ref{Sq}) is obtained, in which Einstein's diffusion coefficient is
slightly modified due to the presence of the quantum position
diffusion coefficient (\ref{coeff}), which depends linearly on the
inverse temperature and vanishes if $\hbar\rightarrow 0$. The
non-Abelian linear Boltzmann equation (\ref{nalbe}) due to its
intrinsic quantum structure should prove as a sound starting point for
the study of quantum kinetic, in which not only the momentum
distribution but also coherence properties are of relevance; it
furthermore has the advantage of being expressed in terms of
quantities of direct physical meaning, such as the dynamic structure
factor and the collision cross-section, for which suitable
phenomenological Ansatz or experimentally determined expressions can
be inserted.
\section*{ACKNOWLEDGMENTS}
The author thanks Prof. L. Lanz for useful discussions and Prof. A.
Barchielli for suggestions. He also thanks Dr.~F.~Belgiorno. This work
was supported by MIUR  under
Cofinanziamento and Progetto Giovani.
  
\end{document}